\documentclass{elsart}

\usepackage{natbib}
\usepackage{amssymb}
\usepackage{graphicx}% Include figure files

\begin{document}

\begin{frontmatter}

\title{Inferring mixed-culture growth from total biomass data in a wavelet approach}

\author[label1]{V. Ibarra-Junquera}\ead{vrani@ipicyt.edu.mx},
\author[label1]{P. Escalante-Minakata}\ead{minakata@ipicyt.edu.mx}, %$\ \ $
\author[label2]{J.S. Murgu\'ia}\ead{ondeleto@uaslp.mx}$\ $ \&
\author[label1]{H.C. Rosu}\ead{hcr@ipicyt.edu.mx}

\address[label1]{IPICyT - Instituto Potosino de
Investigaci\'on Cient\'{\i}fica y Tecnol\'ogica, \footnotesize Apdo
Postal 3-74 Tangamanga, 78231 San Luis Potos\'{\i}, M\'exico}
%\address[label2]{Division of Molecular Biology, IPICyT.}
%\address[label3]{Division of Advanced Materials, IPICyT.}
\address[label2]{%Department of Mathematics and Physical Sciences,
Universidad Aut\'onoma de San Luis Potos\'i, 87545 San Luis
Potos\'{\i}, SLP. M\'exico.}

\bigskip

{\tt Physica A 370 (2006) 777-792}

\begin{abstract}
It is shown that the presence of mixed-culture growth in batch
fermentation processes can be very accurately inferred from total
biomass data by means of the wavelet analysis for singularity
detection. This is accomplished by considering simple
phenomenological models for the mixed growth and the more
complicated case of mixed growth on a mixture of substrates. The
main quantity provided by the wavelet analysis is the H\"older
exponent of the singularity that we determine for our illustrative
examples. The numerical results point to the possibility that
H\"older exponents can be used to characterize the nature of the
mixed-culture growth in batch fermentation processes with potential
industrial applications. Moreover, the analysis of the same data
affected by the common additive Gaussian noise still lead to the
wavelet detection of the singularities although the H\"older
exponent is no longer a useful parameter.
\end{abstract}


\end{frontmatter}

\section{Introduction}

The growth of microbial species in media containing two or several
growth-limiting substrates is of great importance in biotechnology
and bioengineering. The mixed-culture growth occurs in many
industrial processes. A first significant class of such processes is
the traditional fermented foods and beverages in which either
endemic microorganisms or an inoculum with selected microorganisms
are used, see for instance \cite{Szambelan 2004}. Some beverages get
two or more different microorganisms in the inoculum with the
purpose to provide a desired flavor. Evidence of this influence are
presented in the recent paper of \cite{Fleet 2004}, in which the
role of different yeast interactions on the wine flavor is
discussed. However, the phenomenological details and the theory of
the time evolution of the fermentation are as yet poorly understood.
We can also mention the interesting case of the bioethanol
production, in which the substrates used for fermentation typically
consist of a mixture of glucose and fructose. Bioethanol is the
product obtained from the metabolism of microbe mixtures feeding
with this combination of hexoses and pentoses, see e.g., \cite{de
Souza Liberal 2004}. The last relevant example we give is
bioremediation, in which gasoline and chemical spills generally
yield a complex mixture of water-soluble organic compounds. In
gasoline spills, for instance, the four basic compounds are benzene,
toluene, ethylbenzene, and xylene. The consumption of this mixture
by microorganisms is what is defined as the bioremediation process.

In all the aforementioned cases, the presence of different
populations of microorganisms and substrates is a key factor in the
quality and quantity of the final product. Therefore, it is quite
useful to detect the presence or lacking of process of mixed-culture
growth type. Their presence could be used as an estimate of the
right evolution of the process in its early stage. In addition, a
rapid and reasonably accurate test is always useful for saving time
and helping to take quick decisions. It is quite clear then that the
biomass concentration is one of the most needed quantity that should
be measured in fermentation monitoring. The most popular method to
get the biomass concentration is by means of the measurement of the
optical density of centrifugalized samples. However, this procedure
has limited usefulness because it cannot distinguish neither the
living cells from the dead ones, nor the different types of
microorganisms involved in the process. In some cases it is also
possible to correlate the total biomass concentration with the
values of the redox potential of the fermentation.

Recently, new techniques have emerged to quantify the biomass and
distinguish the different microorganisms present in a mixed-culture.
Some of then based on sophisticated equipment (\cite{Callister
2003}, \cite{Madrid 2005} and \cite{Pons 1993}) and others resides
on molecular biology techniques (\cite{de Souza Liberal 2004} and
\cite{Granchi 1999}). All these techniques are very promising in the
study of the dynamics of the mixed-culture growth, although, they
require expensive or complicated procedures. In this paper, we show
that it is possible to infer mixed-culture growth of microorganisms
from their total biomass data, without using such complicated
techniques. The alternative procedure that we put forth here is
based on treating the total biomass data by means of the wavelet
approach for detection of singularities in the growth curves. The
idea is to treat the mixed growth curves as more or less regular
signals that can nevertheless display singularities due to their
compound structure. In the wavelet literature there exist
fundamental papers in which it has been shown that the wavelet
techniques are very efficient in detecting any type of
singularities.

The rest of the paper is organized as follows. In Section 2, we
introduce a simple dynamics of the mixed-growth type and discuss its
basic assumptions. Next, in Section 3, the method of the wavelet
singularity analysis is briefly presented, whereas its application
to the mixed type dynamical curves is enclosed in Section 4. A
conclusion section ends up the paper. An appendix containing the
standard definitions of H\"{o}lder exponents of singularities of
functions is included as well.

\section{A simple mixed-growth model}\label{sec-model}

The technology of batch processes is well developed and numerous
products are obtained in this way. Some products such as food,
beverages, and pharmaceutical ones require precise tracking of the
batch information for safety and regulatory purposes. The primary
objective of monitoring batch processes is to ensure that
significant and sustained changes in the quality of the product
(caused by disturbances and/or faults) are detected as soon as
possible. In that sense, the rapid detection of singularities in the
output of the batch processes offers an interesting solution. The
wavelet analysis for singularity detection is by now well
established but there was no direct application to infer
mixed-growth in the case of batch biochemical processes.

In order to achieve this task we will consider here a fermentation
process consisting of a perfectly stirred tank, where no streams are
fed into it. In the batch fermenter the substrate is converted by
biomass into additional biomass and products. The general
unstructured mass balances for the well-mixed bioreactor can be
represented by the following equations for the concentrations of the
cells and substrates:
%.....................
\begin{eqnarray}
\frac{ \mathrm{d} x_{1,i}}{ \mathrm{d} t}  &=& \  x_{1,i} \  \mu _i\left( x_{2,i} \right) \nonumber\\
\frac{ \mathrm{d} x_{2,i}}{ \mathrm{d} t}  &=& \
-\frac{x_{1,i}}{Y_i}\  \mu _i \left(x_{2,i} \right) \nonumber
\end{eqnarray}
%.....................
where $x_{1,i}$ represent the biomass concentrations, $x_{2,i}$
substrate concentrations %(in $[g/l]$)
and $Y_i$ is the biomass yield, $\mu _i\left( x_{2,i} \right)$ is
the specific growth rate and $i \in \mathbb{R}$ represent the $i$-th
species, allowing for the possibility of multiple kinds of
substrates and microorganisms. The growth rate relates the change in
biomass concentrations to the substrate concentrations. Two types of
relationships for $\mu _i \left( x_{2,i} \right)$ are commonly used:
the substrate saturation model (Monod Equation) and the substrate
inhibition model (Haldane Equation). Both cases will be treated
here. The substrate inhibited growth can be described by
%......................
\begin{eqnarray}
\mu _i\left( x_{2,i}\right)\ = \ \frac{\mu_{max_i}\
x_{2,i}}{K_{{1}_i} + x_{2,i} + K_{{2}_i}x_{2,i}^2}\nonumber
\end{eqnarray}
%........................
where $K_{{1}_i}$ is the saturation (or Monod) constant, $K_2$ is
the inhibition constant and $\mu_{max_i}$ is the maximum specific
growth rate. The value of $K_{1_i}$ expresses the affinity of
biomass for substrate. The Monod growth kinetics can be considered
as a special case of the substrate inhibition kinetics with
$K_{2_i}=0$ when the inhibition term vanishes. For the sake of
simplicity, we will consider only two species and two substrates.
Moreover, we consider that it is possible to measure only the total
biomass concentration. That means that the output of the system
($y$) will be given by
\begin{eqnarray}
y = \ \sum_{i=1}^{m} x_{1,i}\nonumber
\end{eqnarray}
where $m$ is the number of species of microorganisms growing in the
bioreactor (in this work $m=2$). We focus on the following four
cases:
\begin{itemize}
\item[I] The microorganism and substrate concentrations have the same initial
conditions, but different growth rates, one with a Haldane type and
one with a Monod type. In addition, quite different values of the
Monod constant will be taken into account.
\item[II] The microorganism and substrate concentrations have different initial
conditions, but the same growth rates.
\item[III]The microorganism and substrate concentrations have different initial
conditions and different growth rates, one with a Haldane type and
one with a Monod type.
\item[IV] The microorganism and substrate concentrations have the same initial
conditions and the same growth rates, but with different values of
the maximal growth rate.
\end{itemize}

\begin{table}[ht]
\caption{The initial conditions and the values of the employed
parameters of the mixed-growth process model.} \label{tab:1}
\begin{tabular*}{\hsize}{llllllll}\hline
    Symbol     & Meaning                 & Values &        &        &        & Units    &       \\
               &           & \small{Case I}  & \small{Case II}  & \small{Case III} & \small{Case IV} &   & \\\hline
 %              &                         &        &        &        &        &          &       \\ \hline
 $\mu_{max_1}$ & Maximal growth rate     & $1$    & $ 1 $  & $ 1 $  & $0.9$  &  $[l/h]$ &       \\
 $K_{{1}_1}$   & Saturation parameter    & $0.03$ & $0.03$ & $0.03$ & $0.03$ &  $[g/l]$ &       \\
 $K_{{2}_1}$   & Saturation parameter    & $0.5$  & $0.5$  & $0.02$ & $0.5$  &  $[g/l]$ &       \\
 $Y_1$         & Yield coefficient       & $0.5$  & $0.5$  & $0.5$  & $0.5$  &  $\ -$   &       \\
 $x_{{1}_1}^0$ & Initial biomass conc.   & $0.1$  & $0.1$  & $0.1$  & $0.25$ &  $[g/l]$ &       \\
 $x_{{2}_1}^0$ & Initial substrate conc. & $10$   & $10$   & $10$   & $10$   &  $[g/l]$ &       \\
 $\mu_{max_2}$ & Maximal growth rate     & $1$    & $1$    & $1$    & $1$    &  $[l/h]$ &       \\
 $K_{{1}_2}$   & Saturation parameter    & $0.3$  & $0.03$ & $0.03$ & $0.03$ &  $[g/l]$ &       \\
 $K_{{2}_2}$   & Inhibition parameter    & $0$    & $0.5$  & $0$    & $0.5$  &  $[l/g]$ &       \\
 $Y_2$         & Yield coefficient       & $0.5$  & $0.5$  & $0.5$  & $0.5$  &  $\ -$   &       \\
 $x_{{1}_2}^0$ & Initial biomass conc.   & $0.1$  & $0.2$  & $0.1$  & $0.25$ &  $[g/l]$ &       \\
 $x_{{2}_2}^0$ & Initial substrate conc. & $10$   & $ 5 $  & $ 6 $  & $ 10 $ &  $[g/l]$ &       \\ \hline
\end{tabular*}
\end{table}
Table~\ref{tab:1} shows the variables and parameter values used to
simulated the two species growing in the two different substrates,
under the four cases under consideration.

 \newpage

\section{Mixed cultures on mixtures of substrates}

When microbes are grown in a batch reactor containing a surplus of
two substrates, one of the substrates is generally exhausted before
the other, leading to the appearance of two successive exponential
growth phases. This phenomenon could be noticeable at simple view in
the biomass signal or unnoticed due to its nature or due to additive
noise present in the signal. In general, such type of phenomenon is
known as growth of mixed cultures on mixtures of substrates (MCMS).

The growth of MCMS is a phenomenon of practical and theoretical
interest. The fundamental understanding of this problem has impact
on many practical fields such as food processing, production of
ethanol from renewable resources, bioremediation and microbial
ecology, among many others.

To study the usage of the wavelet approach in the detection of MCMS
growth, we consider the recent model proposed by Reeves 2004
\cite{MCMS-model 2004}, which takes into account such type of
growth.

Within this section, the index $i$ will denote the species number,
and the index $j$ will stand for the substrate number. Thus, $c_i$
denotes the concentration of the $i$th species, $s_j$ denotes the
concentration of the $j$th substrate, $e_{ij}$ denotes the
concentration of the lumped system of inducible enzymes catalyzing
the uptake and peripheral catabolism of $s_j$ by $c_i$. Here, $c_i$
and $s_j$ are based on the volume of the chemostat, and expressed in
the units gdw/l and g/l, respectively.  $e_{ij}$ is based on the dry
weight of the biomass, and expressed in the units g/gdw.

\begin{eqnarray}
    r^s_{ij} &=& V^s_{ij}\,e_{ij}\,\frac{s_j}{K^s_{ij}+s_j}\nonumber\\
    r^x_{ij} &=& k^x_{ij}\,x_{ij}\nonumber\\
    r^e_{ij} &=& V^e_{ij}\,\frac{x_{ij}}{K^e_{ij}+x_{ij}}\nonumber\\
    r^{ast}_{ij} &=& k^{ast}_{ij}\nonumber\\
    r^d_{ij} &=& K^d_{ij}\,e_{ij}\nonumber
\end{eqnarray}

\begin{eqnarray}
    \frac{d\,s_j}{d\,t} &=& D \left( s^f_j-s_j \right)- r^s_{1j}\,c_1-r^s_{2j}\,c_2\\
    \frac{d\,e_{ij}}{d\,t} &=& V^e_{ij}\,\frac{e_{ij}\,\sigma_{ij}}{\bar{K}^e_{ij}+e_{ij}\,\sigma_{ij}}+k^{ast}_{ij}-k^d_{ij}\,e_{ij}-r^g_i\,e_{ij}\\
    \frac{d\,c_i}{d\,t} &=& \left( r^g_i -D \right) c_i
\end{eqnarray}
where
\begin{eqnarray}
    \bar{K}^e_{ij} &=&  \frac{K^e_{ij}\,k^x_{ij}}{V^s_{ij}}, \ \ \  \sigma_{ij} \,=\, \frac{s_j}{K^s_{ij}+s_j}\nonumber\\
    r^g_i &=& Y_{i1}\,r^s_{i1}+Y_{i2}\,r^s_{i2}\nonumber
\end{eqnarray}

Reeves et al \cite{MCMS-model 2004} comment that a plausible
experimental situation is the case  of \emph{Escherichia coli} and
\emph{Pseudomonas aeruginosa}, in which, \emph{E. coli} prefers a
sugar over an organic acid, and \emph{P. aeruginosa} prefers the
organic acid over the sugar.

\begin{table}[ht]
\caption{Parameter values used in the MCMS growth model \cite{MCMS-model 2004}} \label{tab:2}
\begin{tabular*}{\hsize}{llllll}\hline
  $V^s_{11}=1000$         & $V^s_{12}=1000$         &   $V^s_{21}=1000$       &   $V^s_{22}=1000$       &  g/g\,h   \\
  $K^s_{11}=0.01$         & $K^s_{12}=0.01$         &   $K^s_{21}=0.01$       &   $K^s_{22}=0.01$       &  g/l      \\
  $V^e_{11}=0.0025$       & $V^e_{12}=0.0020$       &   $V^e_{21}=0.0006$     &   $V^e_{22}=0.0036$     &  g/gdw\,h \\
  $\bar{K}^e_{11}=0.0017$ & $\bar{K}^e_{12}=0.0032$ & $\bar{K}^e_{21}=0.0013$ & $\bar{K}^e_{22}=0.0030$ &  g/gdw    \\
  $k^d_{11}=0.01$         & $k^d_{12}=0.01$         &   $k^d_{21}=0.01$       &   $k^d_{22}=0.01$       &  l/h      \\
  $k^{\ast}_{11}=10^{-2}\,V_{11}$ & $k^{\ast}_{12}=10^{-2}\,V_{12}$ & $k^{\ast}_{21}=10^{-2}\,V_{21}$   &   $k^{\ast}_{22}=10^{-2}\,V_{22}$  &     g/gdw\,h    \\
  $Y_{11}=0.41$           &   $Y_{12}=0.24$         &   $Y_{21}=0.35$         &   $Y_{22}=0.20$         &  g/g & \\ \hline
\end{tabular*}
\end{table}
In order to have the batch regime in the bioreactor we set parameter
$D=0$ and also employ Reeves' parameters $s^{f}_{1}=1$ and
$s^{f}_{2}=2$. Table~\ref{tab:2} shows the rest of the parameter
values used to simulate the growth of the two species on the two
different substrates in the MCMS conditions.

\newpage

\section{Measuring regularity with the wavelet transform}

Let us think of the total biomass of the mixed-growth curves as a
signal. In general, performing the analysis of a signal means to
find the regions of its regular and singular behavior. Usually the
singularities are very specific features for signal
characterization. As it has been pointed in the seminal paper of
\cite{Mallat1}, the regularity of a signal treated as a function can
be characterized by H\"{o}lder exponents. The wavelet transform has
been demonstrated to be a tool exceptionally well suited for the
estimation of H\"{o}lder exponents (for their definitions see the
Appendix).

\subsection{The wavelet transform}

Let $L^2(\mathbb{R})$ denote the space of all square integrable
functions on $\mathbb{R}$. In signal processing terminology,
$L^2(\mathbb{R})$ is the space of functions with finite energy. Let
$\psi(t) \in L^2(\mathbb{R})$ be a fixed function. The function
$\psi(t)$ is said to be a wavelet if and only if its Fourier
transform, $\hat \psi(\omega) = \int e^{i\omega t} \psi(t) dt$,
satisfies
%%%% ===========
  \begin{eqnarray}
   C_{\psi} = \int_{0}^{\infty}  \frac{|\hat\psi(\omega)|^2}{|\omega|} d\omega < \infty.\label{eq-admi}
  \end{eqnarray}
%%%%% ===========
The non-divergent relation given by Eq.~(\ref{eq-admi}) is called
the {\it admissibility condition} in wavelet theory, see for
instance \cite{Ingrid} and \cite{Mallat1}. It implies that the
wavelet must have a zero average on the real line
%%%%% ===========
 \begin{eqnarray}
  \int_{-\infty}^{\infty} \psi(t) dt = \hat \psi(0) = 0,\label{eq-zero}
 \end{eqnarray}
%%%%% ===========
and therefore it must be oscillatory. In other words, $\psi$ must be
a sort of {\it wave}~(\cite{Ingrid, Mallat1}). Based on $\psi(t)$,
one defines the functions $\psi_{a, b}$ as follows
%%%%% ===========
 \begin{eqnarray}
  \psi_{a, b}(t) = \frac{1}{\sqrt{a}} \psi \left( \frac{t - b}{a} \right),\label{eq-waves}
 \end{eqnarray}
%%%%% ===========
where $b \in \mathbb{R}$ is a translation parameter, while $a \in
\mathbb{R}^{+} ~ (a \neq 0)$ is a dilation or scale parameter. The
factor $a^{-1/2}$ is a normalization constant such that $\psi_{a,
b}$ has the same energy for all scales $a$. One notices that the
scale parameter $a$ in Eq.~(\ref{eq-waves}) is a measures of the
dilations of the spatial variable $(t - b)$. In the same way the
factor $a^{-1/2}$ measures the dilations of the values taken by
$\psi$. Because of this, one can decompose a square integrable
function $f(t)$ in terms of the dilated-translated wavelets
$\psi_{a, b}(t)$. We define the wavelet transform (WT) of $f(t)\in
L^2(\mathbb{R})$ by
%%%%% ===========
\begin{eqnarray}
    W_f(a, b) &=& \langle f,\psi_{a,b} \rangle = \int_{-\infty}^{\infty} f(t) \bar{\psi}_{a, b}(t) dt \nonumber\\
              & =& \frac{1}{\sqrt{a}} \int_{-\infty}^{\infty} f(t) \bar{\psi} \left(\frac{t - b}{a} \right) dt,     \label{eq-CWT}
\end{eqnarray}
%%%%% ===========

where $ \langle ~, ~ \rangle $ is the scalar product in $
L^2(\mathbb{R})$ defined as $\langle f, g \rangle := \int f(t)
\bar{g}(t)dt$, and the bar symbol denotes complex conjugation. The
WT given by Eq.~(\ref{eq-CWT}) measures the variation of $f$ in a
neighborhood of size proportional to $a$ centered on point $b$. In
order, to reconstruct $f$ from its wavelet transform (\ref{eq-CWT}),
one needs a reconstruction formula, known as the resolution of the
identity (\cite{Ingrid, Mallat1}).
%%%%% ===========
\begin{eqnarray}
  f(t) = \frac{1}{C_{\psi}} \int_{0}^{\infty} \int_{-\infty}^{\infty} W_f(a, b) \psi_{a, b}(t) \frac{da\ db}{a^2}. \label{inversa}
\end{eqnarray}
%%%%% ===========

From the above equation we can see why the condition given by
Eq.~\ref{eq-admi} should be imposed. One fundamental property that
we require in order to analyze singular behavior is that $\psi(t)$
has enough vanishing moments as argued in the works of \cite{arne1}
and \cite{mallat}. A wavelet is said to have $n$ vanishing moments
  if and only if it satisfies

%%%%% ===========
\begin{eqnarray}
  \int_{-\infty}^{\infty} t^k \psi(t) dx\ &=&\ 0,\ {{\rm for}\ k = 0, 1, \ldots , n - 1}
\end{eqnarray}
and
\begin{eqnarray}
  \int_{-\infty}^{\infty} t^k \psi(t) dt\ &\neq&\ 0, \ {{\rm for}\ k \geq n.} \label{momentos}
\end{eqnarray}
%%%%% ===========

  This means that a wavelet with $n$ vanishing moments is orthogonal to
  polynomials up to order $n - 1$. In fact, the admissibility condition given by
  Eq.~(\ref{eq-admi}) requires at least one vanishing moment. So the wavelet
  transform of $f(t)$ with a wavelet $\psi(t)$ with $n$ vanishing moments
  is nothing but a ``smoothed version'' of the $n$--th derivative of
  $f(t)$ on various scales.   In fact, when someone is interested to
  measure the local regularity of a signal this
  concept is crucial (see for instance \cite{Ingrid,Mallat1}).

\subsection{Wavelet singularity analysis}
%% ============================================

  The local regularity of a function $f$ at a point $t_0$ is often
  measured by its H\"{o}lder exponent. The H\"{o}lder exponent $\alpha$ measures
  the strength of a singularity at a particular point $t_0$, where
  $t_0$ belongs to the domain of $f$, see the Appendix.
  It is important to point out
  that if the singular part of a function $f$ in the
  neighborhood of $t_0$ is of the type
  $|t - t_0|^\alpha$,
  then it corresponds to a {\bf \it cusp} and in this case the singular behavior
  is fully characterized by its
  H\"{o}lder exponent. However, there exists functions that involve oscillating
  singularities which have to be described by an additional quantity: an oscillating
  exponent (\cite{arne2, arne3}). In such a case, the oscillation
  has to be analyzed carefully. Such functions can not be
  fully characterized only by the H\"{o}lder exponent.
  In this work, we will only consider functions whose singularities are not
  oscillating.

  One classical tool to measure the regularity of a function $f(t)$ is to look
  at the asymptotic decay of its Fourier transform $\hat{f}(\omega)$ at infinity.
  However, the Fourier transform
  is not well adapted to measure the local regularity of functions,
  because it is global and provides a description of the overall
  regularity of functions (\cite{mallat,Mallat1}). Consequently, we
  need another way to characterize local signal regularity.

  In the works \cite{arne1, Ingrid, mallat, Mallat1} it is shown that the
  WT provides a way of doing a precise analysis of the regularity
  properties of functions. This is made possible by the scale parameter.
  Due to its ability to focus on singularities in the signals,
  the WT is sometimes referred to as 'mathematical microscope'
  ~(\cite{arne1, Ingrid, mallat, Mallat1}), where the wavelet used determines
  the optics of the microscope and its magnification is given by
  the scale factor $a$.

  The WT modulus maxima (\textsc{WTMM}) decomposition introduced by
  \cite{mallat} provides a local analysis of the singular behavior of signals.
  In the works of Mallat \cite{mallat, Mallat1} it has been shown
  that for cusp singularities the location of the
  singularity can be detected and the related exponent can be recovered from
  the scaling of the WT along the so-called {\em maxima line} (\textsc{Wtmml}
  for short), which is convergent towards
  the singularity. This is a line where the WT reaches local maximum with respect
  to the position coordinate. Connecting such local maxima within the continuous
  WT 'landscape' gives rise to the entire tree of maxima lines.
  Restricting oneself to the collection of such maxima lines provides a particularly
  useful representation of the entire WT. It incorporates the main characteristics
  of the WT: the ability to reveal the {\em hierarchy} of (singular) features,
  including the scaling behavior. \\

An other key concept, in addition to vanishing moments, used to
characterize the regularity of a function in terms of \textsc{Wtmm}
is given next. Suppose that $\psi$ has compact support $[-C, C]$.
The {\it cone of influence} of $\psi$ at point $t_0$ is the set of
points $(a, b)$ in the scale-space plane or domain, such that $t_0$
is in the support of $\psi_{a,b}(t)$. We will denote the
scale--space plane or domain of the WT as the $(a, b)$-plane or the
$(a, b)$-domain. Since the support of $\psi((t - b) / a)$ is $[b -
Ca, b + Ca]$, the point $(a, b)$ belongs to the cone of influence of
$t_0$ if
%%%%% ===========
\begin{eqnarray}
 |b - t_0| \leq Ca. \label{eq-cono}
\end{eqnarray}
%%%%% ===========
The function $f(t)$ has a H\"{o}lder exponent $\alpha \in (k, k+1)$
at $t_0$, if and only if there exists a constant $A>0$ such that at
each modulus maxima $(a,b)$ in the cone defined by
Eq.~(\ref{eq-cono}) one has
%%%%% ===========
\begin{eqnarray}
  |W_f(a, b)| \leq A a^{\alpha + 1/2}, \quad a\to 0, \label{eq-WT-holder}
\end{eqnarray}
%%%%% ===========
(see \cite{mallat, Mallat1}). Here it is assumed that the wavelet
has at least $n>\alpha$ vanishing moments. If $f(t)$ is regular at
$t_0$ or, if the number of vanishing moments is too small, i.e., $n
< \alpha$, one obtains for $a\to 0$ a scaling behavior of the type
%%%%% ===========
\begin{eqnarray}
  |W_f(a, b)| \leq A a^{n + 1/2}. \label{eq-moreR}
\end{eqnarray}
%%%%% ===========
The scaling behavior of the {\sc Wtmml} is given in
Eq.~(\ref{eq-WT-holder}) and can be rewritten as follows
%%%%% ===========
\begin{eqnarray}
    \log |W_f(a,b)|  \leq \log A + \left(\alpha + \frac{1}{2}\right) \log a. \label{WT-singularity}
\end{eqnarray}
%%%%% ===========
The global H\"{o}lder regularity at $t_0$ is thus the maximum slope
$-\frac{1}{2}$ of $\log |W_f(a,b)|$ as a function of $\log a$ along
the maxima line converging to $t_0$. \\

\section{Results and discussion}
 In this section, we present the
results we obtained using the singularity detection procedure
described in the previous section. The signal to be analyzed,
$f(t)=y$, represents the evolution in time of the total biomass
concentration for the fermentation process described in Section
\ref{sec-model} that includes four different cases as specified
therein. In all the wavelet-related calculations we employed as
mother wavelet the first derivative Gaussian $\psi '(t)
=d/dt(e^{-t^2/2})$ having only one vanishing moment. The final goal
is always to calculate the H\"older exponent of the singularities
for such processes because it is a direct measure of the
irregularity of a signal (function) at the singular point $t_0$, in
the sense that higher values of it correspond to more regular
functions than the lower values.

Figure 1 {\it{a,b,c}} shows the performance of the wavelet
singularity analysis as applied to Case I (same initial conditions
but different kinetic rates). We obtain a H\"older exponent of quite
high value.

The following figure shows the performance of the scheme applied
to Case II (same growth rates but different initial conditions).
In this case, the H\"older exponent of the mixed growth
singularity is lower than in Case I.

Similarly to the previous cases, Fig. (3) presents the graphical
results for Case III (different initial conditions and different
growth rates). Although, the singularity looks very mild in the
time evolution of the total biomass concentration the wavelet
analysis is
able to detect it with high precision. % The H\"older exponent

Finally, Case IV (same initial conditions, same growth rates but
with different values of their maximal growth rates) is graphically
analyzed in Fig. (4). For this case we obtained the lowest H\"older
exponent.

Although the latter two cases seem to correspond to almost
overlapping of the WTMML pointing to bifurcation phenomena we are
still not at the threshold of a completely different behavior of the
log plots generated by bifurcations. This could be explained by the
fact that the strength of the first singularity is bigger with
respect to the second one.

\subsection{Wavelet analysis for the MCMS case}

The MCMS case is the most interesting case that we discuss here
because we will show that it is possible to infer in a very
accurate manner by means of WT the moment in which the
microorganisms switch their carbon source. In order to understand
the detailed dynamics of this combined growth, we first apply
separately the WT approach to the two biomass signals $y=c_1$
(Fig. 5) and $y=c_2$ (Fig. 6) and then to the total signal
$y=c_1+c_2$ (Fig. 7).

It is worth noting that the H\"older exponent is bigger than one, a
quite interesting feature which means that the singularity lies in
the second derivative of the biomass signal. This result gives
further opportunities to characterize the nature of the singularity
because it suggests that the type of growth can be inferred from the
order of the derivative in which the singularity occurs directly
given by the value of the H\"older exponent. The latter fact is a
great simplification with respect to the analytical search of the
singularities which implies obtaining the analytical solution of the
given dynamical growth model. Moreover, even in such fortunate
cases, the analytical solutions could be subject to fixed parameter
values of the model. On the other hand, the WT numerical approach
allows the singularity analysis even in the case of time-varying
parameters.

\subsection{Wavelet analysis for the noisy data case}

It is well known that in some cases the amplitude of the Gaussian
noise affecting the on-line signals can be an important annoying
factor. Therefore, a good analysis should be robust in such cases.
Thus, we provide here the WT analysis for the MCMS biomass signals
in the presence of white noise, that is, in the next figures (8-10)
we consider signals of the form $y_{i}=c_i+\epsilon (t)$ and their
sum, where $\epsilon (t)$ stands for the functional form of the
noise.

We notice from the corresponding plots that the noisy data do not
allow to obtain global H\"older exponents in a straightforward
manner, which is a result already reported in the wavelet literature
(\cite{Mallat1} and \cite{noise}). On the other hand, the
singularity detection is robust with respect to the noise for
reasonable levels of its amplitude. In addition, Figure~9 gives us
the hint that when the cones of influence produced by the Gaussian
noise enter the scales of the singularity the cone of the latter
becomes undistinguishable from those of the noise. This remark could
be used as a sort of resolution criterium of the WT method in the
presence of noise. Therefore, one can determine a critical amplitude
of the noise for which the WT approach looses its applicability.

\section{Concluding remarks}

We showed here explicitly how the wavelet singularity analysis can
be applied to infer mixed growth behavior of fermentation processes
using only total biomass data. We prove that the wavelet analysis is
very accurate for all the cases we considered. A very interesting
feature of our research is that the H\"older exponent is sensitive
to the type of the mixed-growth phenomenon, more specifically
depends on the parameters of the growth processes and on their
initial conditions. The MCMS case points to the remarkable
technological possibility of detecting the change of the substrate
uptake since the singularity appears in the second derivatives of
the biomass signal. This can lead one to think of the possibility to
infer substrate contaminations based only in the analysis of the
biomass data. In addition, our results for the noisy data clearly
hint to the fact that the wavelet singularity analysis maintains its
attractive features even in these more difficult but realistic case.
We hope that in future works we could find out the mathematical
relationships implied by this possible correlation. It might allow
the usage of the H\"older exponent as an identification criterium of
the more specific nature of mixed-growth processes.

\section*{Acknowledgements}

This work has been partially supported by CONACyT project 46980 and
the PROMEP fellowship 103.5/03/11118.

%% ==========================
\section*{Appendix} \label{Ap-H}
%% ==========================

  A function $f: \mathbb{R} \to \mathbb{R}$ is said to be H\"{o}lder continuous of exponent
  $\alpha$ ($0 < \alpha < 1$) if, for each bounded interval $(c, d) \subset \mathbb{R}$,
   we can find a positive constant $K$ such that

%%%%% ===========
\begin{eqnarray}
    | f(t) - f(t_0)| \le K |t - t_0|^\alpha\label{HOLDER:1}
\end{eqnarray}
%%%%% ===========

  for all $t, t_0 \in (c, d)$. \\ %, and constant $0< K < \In$. \\

  The space of H\"{o}lder continuous functions is denoted $C^{\alpha}$.
  %and we can make the same definition on $\R^n$.
  A function is said to be
  $C^{n + \alpha}$ if it is in $C^{n}$ and its $n$th %partial
  derivative is H\"{o}lder continuous with exponent $\alpha$.
  %These H\"{o}lder spaces are appropriate for non--linear problems \cite{Falconer}.
  Thus, if we consider the H\"{o}lder exponent %$\a$ in the case when
   $n < \alpha < n + 1$, with $n \in \mathbb{N}$, %we may claim that
  the function can be differentiated $n$ times, but the
  $(n + 1)$th derivative does not exist.
  Therefore, a function with a H\"{o}lder exponent %of strength $\a$ with
  $n < \alpha < n + 1$ is said to be singular in the $n$th derivative.
  Keeping this in mind, let us give the following definition of
  the H\"{o}lder regularity of a function ~\cite{Ingrid, mallat, Mallat1}. \\

%\begin{Definition}[Regularity]
%  \label{regularity}
  \begin{itemize}

   \item []

   \item %[$\bullet$]
    Let $n \in \mathbb{N}$ and $ n \le \alpha < n + 1$. A function
    $f(t)$ has a {\em local} H\"{o}lder exponent $\alpha$ at $t_0$ if and only
    if there exist a constant $K > 0$, and a polynomial
    $P_n(t)$ of order $n$, such that

%%%%% ===========
\begin{eqnarray}
\forall t \in \mathbb{R}, \qquad   | f(t) - P_n(t - t_0)| \le K |t -
t_0|^{\alpha} \label{holder}
\end{eqnarray}
%%%%% ===========

   \item%[$\bullet$]
    The function $f(t)$ has a {\em global} H\"{o}lder exponent
    $\alpha$ on the interval $(c, d)$ if and only if there is a constant $K$ and a
    polynomial of order $n$, $P_n(t)$, such that equation (\ref{holder}) is
    satisfied for all $t\in (c, d)$.

   \item %[$\bullet$]
   The H\"{o}lder {\em regularity} of $f(t)$ at $t_0$ is the %superior bound
   supremum of the $\alpha$ such that $f(t)$ is H\"{o}lder $\alpha$ at $t_0$.

   \item %[$\bullet$]
   The $n$th derivative of a function $f(t)$ is {\em singular} at
     $t_0$ if $f(t)$ has a local H\"{o}lder exponent $\alpha$ at $t_0$ with
     $n < \alpha < n + 1$.
   \\

  \end{itemize}

%\end{Definition}

  A function $f(t)$ that is continuously differentiable at a given point has a %the
  H\"{o}lder exponent not less than 1 at this point.
  If $\alpha \in (n , n+1)$ in (\ref{holder}) then
  $f(t)$ is $n$ times but not $(n + 1)$ times differentiable at the point $t_0$, and
  the polynomial $P_n(t)$ corresponds to the first $(n + 1)$ terms of the Taylor series
  of $f(t)$ around $t = t_0$.
  For example, if $n = 0$, we have
  $P_0(t - t_0) = f(t_0)$.

  \newpage

  \begin{center}
  \section*{Figure Captions}
  \end{center}

  {\em Fig. 1}

 {\it a}) The time evolution of the total biomass concentration signal for Case I. {\it b}) The wavelet cones of influence
  corresponding to this case showing a very accurate identification of the two singularity points presented in the signal, of which the
  first one allows to infer the presence of the mixed growth feature of the fermentation process whereas the second one is associated with
  the end of the fermentation batch cycle.\\ \noindent {\it c}) From the slope in the double logarithmic plot, the
H\"older coefficient of the mixed
  growth singularity
  is calculated as $\alpha =0.95$.

\medskip

{\em Fig. 2}

{\it a}) The time evolution of the total biomass concentration
signal for Case II.\\ \noindent {\it b}) The wavelet cones of
influence
  corresponding to this case again showing the accurate identification of the two singularity points, of the same type,
  respectively, as in Fig.~1.
\\ \noindent
{\it c}) The H\"older coefficient of the mixed growth singularity
  is now $\alpha =0.88$.

\medskip

{\em Fig. 3}

Same caption comments as in the previous figures but for Case III.
  %{\it a}) The time evolution of the total biomass concentration signal for Case III, {\bf b)} The wavelet cones of influence
  %corresponding to this case again showing the accurate identification of the two singularity points, of the same type,
  %respectively, as in Fig.~1, {\bf c)}
  The H\"older coefficient of the mixed growth singularity
  is now $\alpha =0.92$.

\medskip

{\em Fig. 4}

 Same caption comments as in the previous figures but for Case IV.
  The value of the he H\"older coefficient for the mixed growth singularity
  is $\alpha =0.84$.

 \medskip

 {\em Fig. 5}

  Same caption comments as in the previous figures but for the MCMS biomass signal $y=c_1$.
  %{\bf a)} The time evolution of the total biomass concentration signal for Case MCMS $c_1$, {\bf b)} The wavelet cones of influence
  %corresponding to this case again showing the accurate identification of the two singularity points, of the same type,
  %respectively, as in Fig.~1, {\bf c)}
  The H\"older coefficient of the mixed growth singularity
  is now $\alpha =1.89$.

\medskip

{\em Fig. 6}

 %%%%%%%%%%%%%%%%%%   FIGURA 6

Same caption as in the previous figures but for the MCMS biomass signal $y=c_2$. %but for case III.
  %{\bf a)} The time evolution of the total biomass concentration signal for Case MCMS $c_2$, {\bf b)} The wavelet cones of influence
  %corresponding to this case again showing the accurate identification of the two singularity points, of the same type,
  %respectively, as in Fig.~1, {\bf c)}
  The H\"older coefficient of the mixed growth singularity
  is now $\alpha =1.87$.

\medskip

\newpage

{\em Fig. 7}

Same caption comments as in the previous figures but for the total
MCMS signal.
  %{\bf a)} The time evolution of the total biomass concentration signal for total MCMS signal, {\bf b)} The wavelet cones of influence
  %corresponding to this case again showing the accurate identification of the two singularity points, of the same type,
  %respectively, as in Fig.~1, {\bf c)}
  The H\"older coefficient of the mixed growth singularity
  is now $\alpha =1.88$.

\medskip

{\em Fig. 8}

MCMS data corresponding to $c_1$ with a small amplitude Gaussian
noise added.
  %{\bf a)} The time evolution of the total biomass concentration signal for Case III, {\bf b)} The wavelet cones of influence
  %corresponding to this case again showing the accurate identification of the two singularity points, of the same type,
  %respectively, as in Fig.~1, {\bf c)}
  From the bottom plot {\it c}) one can see that because the curve is not a straight line one cannot get a global H\"older
  coefficient from its slope.

\medskip

{\em Fig. 9}

MCMS data corresponding to $c_2$ with a small amplitude Gaussian
noise added.
  %{\bf a)} The time evolution of the total biomass concentration signal for Case III, {\bf b)} The wavelet cones of influence
  %corresponding to this case again showing the accurate identification of the two singularity points, of the same type,
  %respectively, as in Fig.~1, {\bf c)}
  The H\"older coefficient is not a useful concept in this case. %of the mixed growth singularity
  %is now $\alpha =0.92$.

\medskip

{\em Fig. 10}

   %%%%%%%   FIG. 10
MCMS data corresponding to the sum $c_1+c_2$ with the same
Gaussian noise added.
  %{\bf a)} The time evolution of the total biomass concentration signal for Case III, {\bf b)} The wavelet cones of influence
  %corresponding to this case again showing the accurate identification of the two singularity points, of the same type,
  %respectively, as in Fig.~1, {\bf c)}
  The concept of global H\"older coefficient is again not useful.

\end{document}